\documentclass[onecolumn,notitlepage,floatfix,amsmath,amssymb,longbibliography]{revtex4-1}

\usepackage{graphicx}
\usepackage{dcolumn}
\usepackage{color}
\usepackage[normalem]{ulem}
\usepackage{amssymb}
\usepackage[toc,page]{appendix}
\newcommand{\be}{\begin{equation}}
\newcommand{\ee}{\end{equation}}

\usepackage{graphicx}
\usepackage{enumerate}

\newcommand{\bom}{\mbox{\boldmath $\omega$}}

\newcommand{\RN}[1]{\textup{\uppercase\expandafter{\romannumeral#1}}}

\begin{document}
\title{Invariant manifolds in stratified turbulence}

\author{N.E. Sujovolsky, G.B. Mindlin, and P.D. Mininni}
\affiliation{
  Universidad de Buenos Aires, Facultad de Ciencias Exactas y
  Naturales, Departamento de F\'\i sica, \& IFIBA, CONICET, Ciudad
  Universitaria, Buenos Aires 1428, Argentina.}

\begin{abstract}
We present a reduced system of 7 ordinary differential equations that
captures the time evolution of spatial gradients of the velocity and
the temperature in fluid elements of stratified turbulent flows. We
show the existence of invariant manifolds (further reducing the system
dimensionality), and compare the results with data stemming from
direct numerical simulations of the full incompressible Boussinesq
equations in the stably stratified case. Numerical results accumulate
over the invariant manifolds of the reduced system, indicating the
system lives at the brink of an instability. Finally, we study the
stability of the reduced system, and show that it is compatible with
recent observations in stratified turbulence of non-monotonic
dependence of intermittency with stratification.
\end{abstract}
\maketitle

%%%%%%%%%%%%%%%%%%%%%%%%%%
\section{Introduction}
%%%%%%%%%%%%%%%%%%%%%%%%%%

Turbulent flows have a huge number of degrees of
freedom (d.o.f.~$\sim \mathrm{Re}^{9/4}$, with Reynolds number
$\mathrm{Re} \gtrsim 10^5$), and are difficult to study
as they are non-Gaussian and intermittent \cite{li_origin_2005,
  xu_flightcrash_2014, international_collaboration_2008}. 
However, important computational and theoretical evidence
indicates that in some flows displaying complex dynamics, there is
strong similarity between long time evolution and solutions of
finite dimensional dynamical systems. 
Historically, attempts
to study complex flows using finite dimensional dynamical
systems first focused on truncating the infinite set of equations
ruling the amplitudes of modes compatible with the boundary
conditions \cite{lorenz63}. A more careful analysis demanded the
separation between central and slaved modes, a hierarchical set of
structures capable of describing the perturbations of a simple
solution. The finite dimensional dynamical systems emerged as the
normal forms on the manifold spanned by the central
modes \cite{gucken83}. These ideas had a global analog in those cases  
where coherent structures could be identified
\cite{holmesberkooz96}. In this case, modes participating of the low
dimensional dynamics are found from the data, by statistical 
methods as the proper orthogonal decomposition. 
%\DEL{(or proper orthogonal decomposition)}.
In these approaches the dissipative nature of the problem also reduces
the d.o.f., and allows to  study a dynamical
system ruling the behavior of active structures, whose attractor
provides a good approximation to the global attractor of the original
problem.

But this is not the only way in which finite dimensional dynamical
systems can help us unveil the dynamics of complex flows.
Extreme events, associated to intermittency, are
important in the atmosphere and the ocean, where turbulence is
also ubiquitous \cite{pearson_log-normal_2018, rorai_turbulence_2014,
  feraco_vertical_2018}. In homogeneous and isotropic turbulence
(HIT) intermittency has been captured by two
dimensional restricted Euler models \cite{meneveau_lagrangian_2011}. 
In this case, the dynamical system is in
material derivatives. Is it possible to find low dimensional dynamical
systems ruling part of the dynamics of more complex flows? 
Previous studies indicate such reduced models are non-viable for
  anisotropic flows \cite{girimaji_1995}. This case is important as 
%\DEL{for example} 
in geophysical scenarios flows are anisotropic and display
internal waves that propagate and interact with the turbulence
%\DEL{and have an even larger number of d.o.f.} 
\cite{pouquet_scaling_2018, maffioli_mixing_2016,
  waite_stratified_2004, riley_fluid_2000}.

Here we derive, with minimal assumptions, a 7 dimensional dynamical
system for the material derivatives of velocity and temperature
gradients in fluid elements of stratified flows, such as those 
in the atmosphere and the ocean. These flows are disordered and
strongly anisotropic, as the evolution of the velocity is
coupled with temperature fluctuations, giving rise to higher
complexity than in HIT. We identify the fixed points of the reduced
system, and invariant manifolds in their
vicinity. We perform direct numerical simulations (DNSs) of stably
stratified turbulence using the full partial differential
equations (PDEs), and show that the system evolves preferentially in
the vicinity of these manifolds. Moreover, many fluid elements 
lay near a manifold corresponding to the boundary of the
convective vertical instability, living at the
brink of convection.

%%%%%%%%%%%%%%%%%%%%%%%%%%%%%%%%%%
\section{The restricted Euler model and its invariant manifolds}
%%%%%%%%%%%%%%%%%%%%%%%%%%%%%%%%%%

%%%%%%%%%%%%%%%%%%%%%%%%%%%%%%%%%%%%%%%%%%
%% 		FIGURE 1		%%
%%%%%%%%%%%%%%%%%%%%%%%%%%%%%%%%%%%%%%%%%%
\begin{figure}
\includegraphics[width=17.9cm]{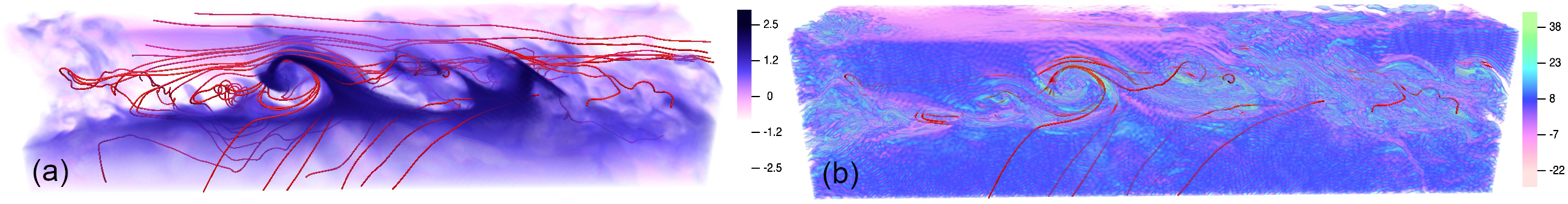}   
\caption{(a) Rendering of the 
  buoyancy $\theta$, in a small subvolume with local convection, in a
  simulation of stably stratified turbulence with $N=8$ using
  $2048\times2048\times256$ grid points ($\approx 1/10$ of the 
  length of the domain is
  shown). Note the entrainment and mixing between denser and 
  lighter fluid in the region with rolls. (b) Rendering of
  $S= \partial_{z} \theta$. Unstable regions have
  $S>N$ (light green), but many points have $S \approx N$
  (blue). In both panels, a few particle trajectories are indicated in
  red.}
\label{f:CONV}
\end{figure}
%%%%%%%%%%%%%%%%%%%%%%%%%%%%%%%%%%%%%%%%%%

We consider the Lagrangian dynamics of an incompressible stratified
flow under the Boussinesq approximation. For a linear density profile,
the Boussinesq equations for the Eulerian velocity ${\bf u}$ and the
buoyancy $\theta$ (proportional to temperature or density
fluctuations) are
\begin{eqnarray}
\label{eq:n-s_strat}
\partial_t {\bf u} +{\bf u}\cdot{\bf \nabla}{\bf
  u} &=& -{\bf \nabla}p - N \theta {\hat z} +\nu \nabla^{2}{\bf u}+{\bf
  f}, \\
\label{eq:theta}
\partial_t \theta+ {\bf u}\cdot{\bf \nabla} \theta
  &=& N {\bf u} \cdot {\hat z}  + \kappa \nabla^{2} \theta,
\end{eqnarray}
where $\nabla \cdot {\bf u}=0$, $p$ is the correction to the
hydrostatic pressure, $N$ the Brunt-V\"{a}is\"{a}l\"{a} frequency
(a non-decreasing function of the density profile steepness, which
sets the level of stratification), $\nu$ the kinematic
viscosity, $\kappa$ the diffusivity, and ${\bf f}$ an external
mechanical forcing. Equations (\ref{eq:n-s_strat}) and
(\ref{eq:theta}) have two controlling dimensionless parameters, the
Reynolds number $\mathrm{Re}=UL/\nu$, and the Froude number
$\mathrm{Fr}=U/(NL)$, where $U$ and $L$ are characteristic velocities
and lengths. While Re controls the strength of the
nonlinearities, Fr measures stratification, with typical
geophysical values of $\textrm{Fr}\approx 10^{-2}$.
Another important parameter is
the gradient Richardson number 
$\mathrm{Ri}_{g} = N(N-\partial_{z}\theta) / (\partial_{z}U_{\perp})^{2}$,
where $U_{\perp}$ is the horizontal velocity \cite{rosenberg_evidence_2015}. 
Pointwise, this number measures the flow vertical stability: When
$\mathrm{Ri}_{g} \leq 1/4$ the flow can undergo shear instabilities
\cite{billant_theoretical_2000}, while for $\mathrm{Ri}_{g}\leq0$ the vertical
buoyancy gradient $\partial_z\theta$ can overcome the
background gradient (controlled by $N$) and local convection can
develop, increasing 
%vertical 
mixing \cite{mashayek}. Figure~\ref{f:CONV} shows as an illustration a
rendering of 
$\theta$ and of $\partial_z\theta$ in a subvolume of a DNS of 
stably stratified turbulence with $N=8$.

From these equations we derive a closed system of Lagrangian equations
for the velocity and buoyancy gradients, for the ideal unforced
case ($\nu = \kappa = {\bf f} = 0$). These provide an
approximation of the fields surrounding an observer moving with the
fluid.
% ADD %%%%%%%%%%%
  We first define $A_{ij}=\partial_{j}u_{i}$ and
  $\theta_{j}=\partial_{j}\theta$ (for $i$ and $j=\{x,y,z\}$), and
  then compute the spatial derivatives of Eqs.~(\ref{eq:n-s_strat})
  and (\ref{eq:theta}) to obtain 
\begin{equation}
\dfrac{DA_{ij}}{Dt} + A_{kj}A_{ik} = - \dfrac{\partial^{2}p}{\partial
  x_{i}\partial x_{j}} - N \theta_{j} \delta_{iz},
\,\,\,\,\,\,\,\,\,\,\,\,\,\,\,\,
\dfrac{D\theta_{j}}{Dt} + A_{kj} \theta_{k} = N A_{zj},
\label{eq:Aij}
\end{equation}
  where $\delta_{ij}$ is the Kronecker delta, and $D/Dt$ is the
  material derivative. We can remove some of the derivatives of the
  pressure in Eqs.~(\ref{eq:Aij}) using the incompressibility
  condition ${\boldsymbol \nabla} \cdot {\bf u} = A_{ii}=0$, which
  from the equation for $A_{ij}$ in Eqs.~(\ref{eq:Aij}) implies 
  $A_{kl}A_{lk} = - \partial_{l}\partial_{l}p - N \theta_{z}$. The
  remaining spatial derivatives of the pressure can be written using
  the pressure Hessian, 
  $H_{ij}=-\partial_{i}\partial_{j} p +
  \delta_{ij} \partial_{k}\partial_{k}p/3$.
  Using these three relations, the equation for $A_{ij}$ can
  be rewritten as
\begin{equation}
\dfrac{DA_{ij}}{Dt} + A_{kj}A_{ik} -\dfrac{\delta_{ij}}{3}
  A_{kl}A_{lk} =  H_{ij} -  N \theta_{j} \delta_{iz} + N \theta_{z}
  \dfrac{\delta_{ij}}{3}.
\label{eq:AijH0} 
\end{equation}
  In restricted Euler models it is often assumed that 
  $H_{ij}\approx 0$, a condition that is not satisfied by the Euler
  equation even in the isotropic case. Interestingly, for stratified
  flows we verified in our DNSs that the diagonal terms of $H_{ij}$
  are small, while off-diagonal terms satisfy on the average the
  relation $|H_{xy}| < |H_{xz}| \approx |H_{yz}|$ and decrease in
  amplitude with increasing $N$ (becoming smaller, albeit of the same
  order than the other terms in the equations, for $N\approx 8$). 
  We next define the scalars
  $Q=-A_{ij}A_{ji}/2$, 
  $R=-A_{ij}A_{jk}A_{ki}/3$, 
  $R_{\theta} = \theta_{i}A_{ij}A_{jz}$, 
  $T= \theta_{i}A_{iz}$, 
  $B= A_{zi}A_{iz}$,
  $A=A_{zz}$, and
  $S=\theta_{z}$. 
  The reduced system of ODEs can thus be obtained by setting
  $H_{ij}\approx 0$, multiplying the equations by $A_{ij}$ and
  $\theta_j$, and using the following relations: 
  $A_{ik}A_{kl}A_{lj} = - Q A_{ij} - R \delta_{ij}$, 
  $A_{ik}A_{kl}A_{lz} = - Q A_{iz} - R \delta_{iz}$, and 
  $A_{zk}A_{kl}A_{lz} = - Q A_{zz} - R$.
  These relations follow directly from the Cayley-Hamilton theorem  
  and the incompressibility condition 
  \cite{vieillefosse_local_1982, cantwell_exact_1992,
    meneveau_lagrangian_2011}. Finally, we obtain
\begin{equation}
\begin{split}
\label{eq:ODEs}
D_{t}{Q}&=-3R+NT , \,\,\,\,\,\,\,\,\,\,
D_{t}{R}=2Q^{2}/3+2NSQ/3+NR_{\theta} , \,\,\,\,\,\,\,\,\,\,
D_{t}{R_{\theta}}= 5QT/3 + 3RS- 4NST/3 - NQA -NR , \\
D_{t}{B}&= 2QA/3+2R-NAS/3-NT ,  \,\,\,\,\,\,\,\,\,\,
D_{t}{T}=-2R_{\theta} - 2SQ/3 + NB -2NS^{2}/3 , \\
D_{t}{A}&=-B-2Q/3- 2NS/3, \,\,\,\,\,\,\,\,\,\,
D_{t}{S}=NA-T .
\end{split}
\end{equation}
These equations give the time evolution
of field gradients moving along fluid trajectories,
and in that frame are a closed set of
7 ordinary differential equations (ODEs). Except for the assumption
that $H_{ij}\approx0$, these ODEs are
exact. The drop of 
%\DEL{the pressure gradients} 
$H_{ij}$ results in a blow up at finite time: eventually,
gradients in the system diverge. However, as we will see, the dynamics
of this system before the divergence will provide significant
information on the full set of PDEs.

%%%%%%%%%%%%%%%%%%%%%%%%%%%%%%%%%%%%%%%%%%%%%%
%%\section{Invariant manifold reduction S=0}%%
%%%%%%%%%%%%%%%%%%%%%%%%%%%%%%%%%%%%%%%%%%%%%%

The system of Eqs.~(\ref{eq:ODEs}) has two sets of fixed points, 
\begin{equation}
\begin{split}
\RN{1}:\ &Q=R=R_{\theta}=T=B=A=S=0  \\
\RN{2}:\ &R_{\theta}=(2N^{2}Q-6Q^{2})/(3N), \ B=2N^{2}/3-2Q, \ 
  T=3R/N, \ A=3R/N^{2},  \ S=2Q/N-N, \
  Q \textrm{ and } R \textrm{ free.}
\end{split}
\end{equation}
Several complex eigenvalues near $\RN{1}$ represent oscillations
(between $B$ and $T$, $R$ and $R_\theta$, and two linear combinations of other
variables), associated, in the stably stratified case, to the effect 
of gravity waves over the gradients.

For $N=0$ (no stratification), the system recovers a central invariant
manifold of restricted Euler models of HIT, the 
Vieillefosse tail \cite{vieillefosse_local_1982, cantwell_exact_1992,
  meneveau_lagrangian_2011}, as
$D_t(4Q^3/27+R^2)= 3N(9NRB+6NRQ+2Q^2T)/2$.
More generally,
$T=NA$, 
$S=N$, and 
$R_{\theta}=NB$ 
is a central invariant manifold valid for all orders in the
nonlinearity, as 
\begin{equation}
\Sigma_0: D_{t}(T-NA)=0, D_{t}S=0, D_{t}(R_{\theta}-NB)=0.
\end{equation}
In this manifold, the dynamics reduces to a 4 dimensional dynamical
system:
$D_{t}{Q}=-3R+NT$, $D_{t}{R}=N^{2} B + 2N^{2} Q/3 +  2Q^{2}/3$,
$D_{t}{B}= -4N^{2}T/3 + 2NR + 2QT/3$, and
$D_{t}{T}=-NB-2N^{3}3-2NQ/3$.
This manifold has a fundamental physical interpretation:
when $S=\partial_{z} \theta=N$ the gradient Richardson number becomes
$\textrm{Ri}_g=0$, and the local buoyancy 
gradient cancels the
background stratification. In the stably stratified case, fluid
elements in this manifold are at the brink of the local convective
instability. They can undergo a sudden and
intermittent convection process \cite{rorai_turbulence_2014},
enhancing vertical dispersion in flows characterized by low
vertical transport \cite{lindborg_vertical_2008,
  aartrijk_single-particle_2008}. 

For the ODEs in Eqs.~(\ref{eq:ODEs}) we can compute local invariant
manifolds, valid in the vicinity of the fixed points (i.e., up to
linear order). Near fixed point $\RN{1}$,
\begin{equation}
\Sigma_{\RN{1}}: D_{t}\left( R_{\theta}/N-B-Q \right) = 0 .
\end{equation}
Fixed points $\RN{2}$ have two other invariant manifolds in their
vicinity
\begin{equation}
\begin{split}
\Sigma_{\RN{2},a}: \ &Q=N^{2}/3, \ \ R \textrm{ free}, \ \ 
  D_{t}(-4R+NT+N^{2}A)=0, \\
\Sigma_{\RN{2},b}: \ &Q=-N^{2}/3, \ \ R \textrm{ free}, \ \  
  D_{t}(R_{\theta}-NB + 8NQ/3)=0.
\end{split}
\end{equation}
As these manifolds require $Q \sim N^{2}$, they are hard
to access and will not play a
relevant role in the dynamics.

%%%%%%%%%%%%%%%%%%%%%%%%%%%%%%%%%%%%%%%%%%
%% 		FIGURE 2		%%
%%%%%%%%%%%%%%%%%%%%%%%%%%%%%%%%%%%%%%%%%%
\begin{figure}
\includegraphics[width=8.9cm]{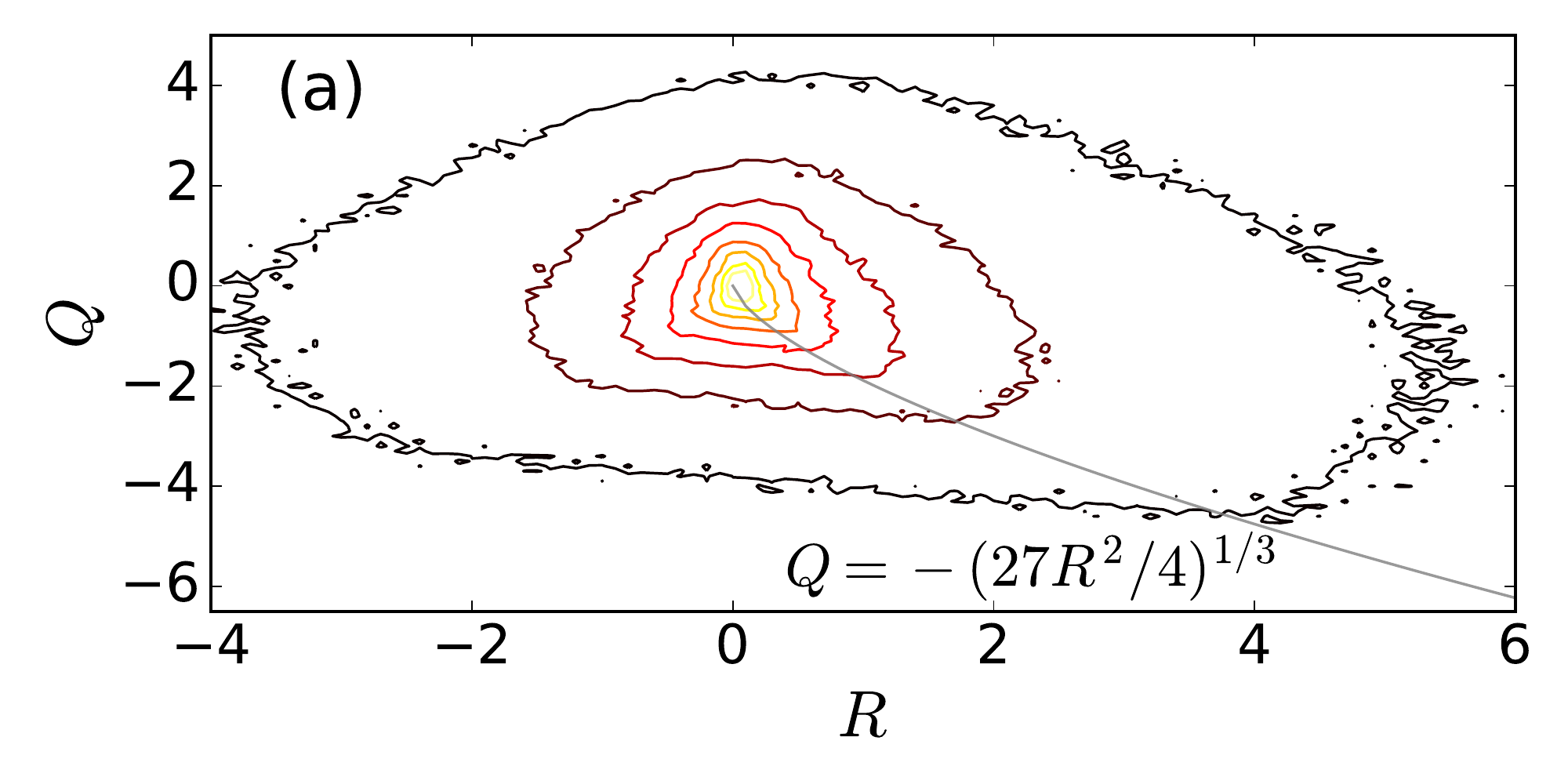}   
\includegraphics[width=8.9cm]{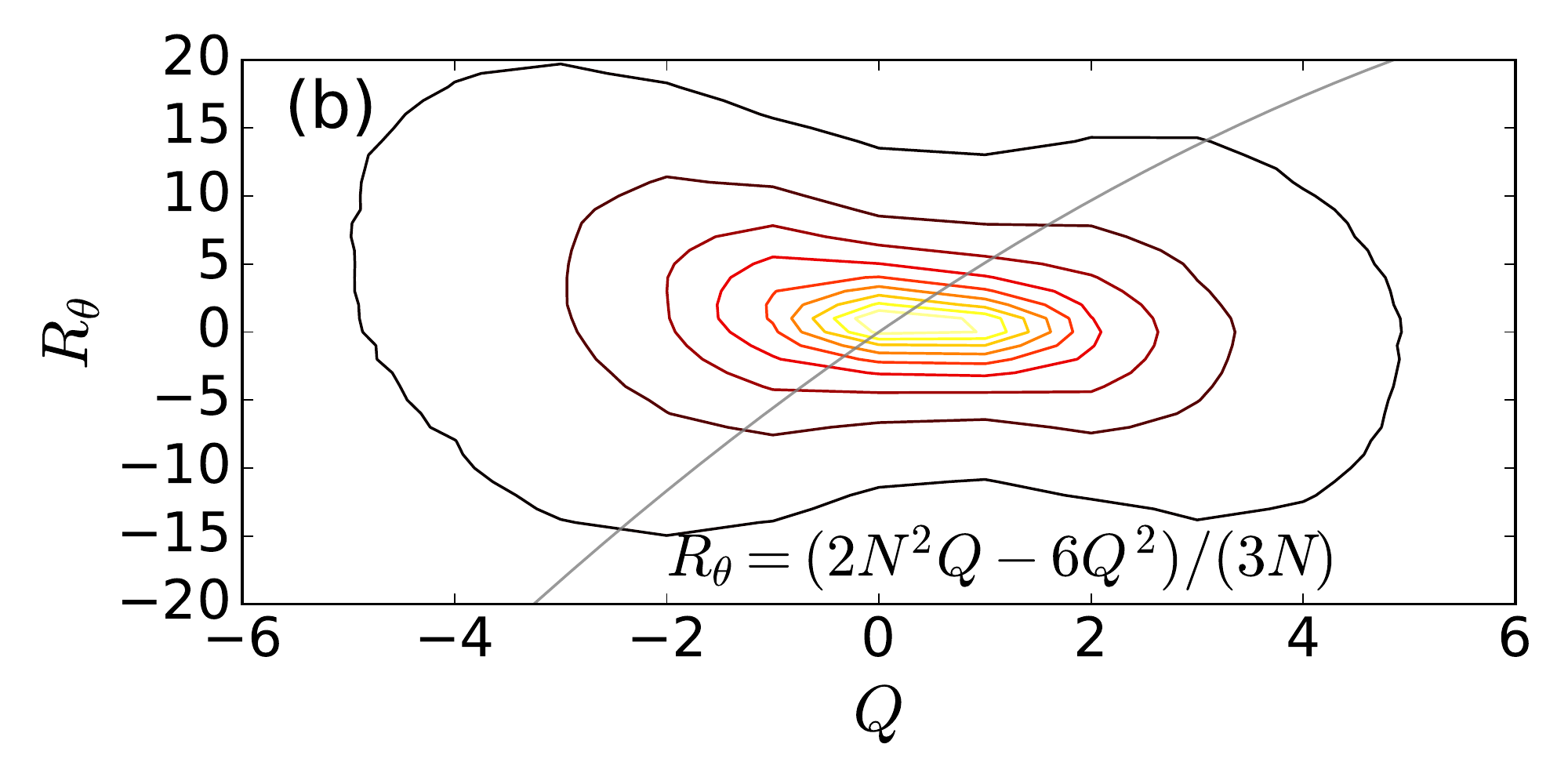}   
\caption{(a) Isocontours of the joint PDF of $R$
  and $Q$, for the DNS with $N=8$ and TG forcing. The gray curve
  indicates $Q=-(27R^{2}/4)^{1/3}$. (b) Same for $Q$ and
  $R_{\theta}$. The gray curve indicates
  $R_{\theta}=(2N^{2}Q-6Q^{2})/(3N)$.}
\label{f:RQ_8}
\end{figure}
%%%%%%%%%%%%%%%%%%%%%%%%%%%%%%%%%%%%%%%%%%

%%%%%%%%%%%%%%%%%%%%%%%%%%%%%%%%%%%%%%%%%%%%
\section{Direct numerical simulations}
%%%%%%%%%%%%%%%%%%%%%%%%%%%%%%%%%%%%%%%%%%%%

%\ADDA{
To study the role of these manifolds we performed DNSs of the full
system of PDEs in Eqs.~(\ref{eq:n-s_strat}) and (\ref{eq:theta}) in
the three-dimensional (3D) turbulent stably stratified case, using
a parallel pseudo-spectral code with a second order Runge-Kutta scheme
in time \cite{GHOST}. Although the reduced 
  ODEs were derived for ${\bf f}=\nu=\kappa=0$, dissipation and
  forcing are needed in the DNSs to ensure numerical stability and to
  compensate for energy dissipation. Thus, turbulence was sustained 
with the mechanical forcing ${\bf f}$, which was set to either
Taylor-Green (TG) forcing \cite{riley_03, sujovolsky_generation_2018} 
(which excites a large-scale horizontal circulation), or to 3D random
(RND) isotropic forcing \cite{Clark} (which excites internal gravity
waves and horizontal winds). The forcing was applied at a wave number
$k_f=1$ for TG forcing and at $k_f=4$ for RND forcing. The
Brunt-V\"{a}is\"{a}l\"{a} frequency $N$ was varied from $8$ to $12$
(in dimensionless units). Elongated three-dimensional periodic domains,
with two different aspect ratios, were used; in all cases, the length
of the domain in the $x$ and $y$ directions was fixed to
$L_x=L_y=2\pi$ (resulting in a domain height $L_z=2\pi/4$ or $2\pi/8$,
depending on the domain aspect ratio), and the grid was 
always isotropic with spatial resolution 
$\Delta = \Delta_x = \Delta_y = \Delta_z$.
The dimensionless controlling parameters were the Froude number
$\textrm{Fr}$ and the Reynolds number $\textrm{Re}$, based on the flow
integral scale $L$ and on the flow r.m.s.~velocity $U$ of order
unity. From these two controlling parameters it is also possible to
define the buoyancy Reynolds number 
$\textrm{Rb}  =   \textrm{Re} \, \textrm{Fr}^{2},$ 
which measures of how strong turbulence is at the buoyancy length
scale. In all simulations, $\textrm{Rb}>1$.
As the flows are anisotropic, several characteristic length scales can
be defined. The flow integral scale (or flow correlation length) in
the parallel direction $L_\parallel$ is close to the isotropic
integral scale $L$, while the perpendicular integral scale $L_\perp$
in all runs is larger. Stratified flows have two other characteristic
length scales associated to the stratification, the buoyancy length
scale $L_b=U/N$, and the Ozmidov length scale $L_{oz}=2\pi/k_{oz}$,
where $k_{oz} = (\epsilon N^{-3})^{-1/2}$ is the Ozmidov wave number
and $\epsilon$ is the energy injection rate. Below the Ozmidov scale,
the flow is expected to slowly recover isotropy. All these parameters
are listed for our runs in table \ref{tab:param}. In each simulation
we integrated ${\cal O}(10^6)$ Lagrangian particles, and stored
the components of the velocity and buoyancy tensor gradients $A_{ij}$
and $\theta_j$ along their trajectories for over 10 turnover times. In
the following, figures show results for the simulation with TG forcing, 
$N=8$, spatial resolution of $768\times768\times192$ grid
points, and aspect ratio 4:4:1; results for other simulations are
qualitatively similar with quantitative differences discussed in the
text.

\begin{table}[b]
\centering
\setlength{\tabcolsep}{.26 cm}
\begin{tabular*}{\textwidth}{p{1.0cm} c c c c c c c c c c c c c c}
\hline \hline
Forcing & $k_{f} $ & $N$ & Aspect ratio & $n_{x}=n_{y}$ & $n_{z}$ &
    $\Delta$ &  $\textrm{Fr}$ & $\textrm{Re}$ & $\textrm{Rb}$ & $L$ &
    $L_{\perp}$ & $L_{\parallel}$ & $L_{b}$ & $L_{oz}$ \\
\hline
TG  & $1$ &$8$  &8:8:1 & 2048 & 256 & $0.003$ &
    $0.03$ &$ 35000$  & $30$ & $0.6$ & 
    $4.9$  & $0.6$ & $0.16$ & $0.18$ \\
TG  & $1$ &$4$  &4:4:1 & 768  & 192   & $0.008$&
    $0.05$ &$ 10000$  & $25$ & $1.2$ & 
    $4.7$  & $1.2$ & $0.24$ & $0.36$ \\
TG  & $1$ &$8$  &4:4:1 & 768  & 192 &  $0.008$ &
    $0.03$ &$ 14000$  & $13$ & $1.1$ & 
    $5.2$  & $1.1$ & $0.15$ &  $0.14$ \\
TG  & $1$ &$12$ &4:4:1 & 768  & 192  &  $0.008$&
    $0.02$ & $15000$  &  $4$ & $0.9$ & 
    $5.5$  & $0.9$& $0.1$   &  $0.07$ \\
RND & $4$ &$8$  &4:4:1 & 768  & 192  &  $0.008$&
    $0.1$  &$ 3000 $  & $36$ & $1.5$ &
    $1.0$ & $1.5$ &  $1$ & $0.17$ \\
\hline
\end{tabular*}
\caption{Relevant parameters of the simulations: Forcing indicates
  the forcing function (either TG or RND), $N$ is the
  Brunt-V\"{a}is\"{a}l\"{a} frequency, the aspect ratio indicates the
  domain aspect ratio in the three spatial directions, $n_{x}=n_{y}$
  and $n_{z}$ are the number of grid points in each direction, and
  $\Delta$ is the grid spatial resolution. $\textrm{Fr}$ is the Froude
  number, $\textrm{Re}$ is the Reynolds number, and $\textrm{Rb}$ is
  the buoyancy Reynolds number. Finally, $L$ is the flow isotropic
  integral scale, $L_{\perp}$ is the perpendicular integral scale,
  $L_b$ is the buoyancy length, and  $L_{oz}$ the Ozmidov length.}
\label{tab:param}
\end{table}

%%%%%%%%%%%%%%%%%%%%%%%%%%%%%%%%%%%%%%%%%%
%% 		FIGURE 3		%%
%%%%%%%%%%%%%%%%%%%%%%%%%%%%%%%%%%%%%%%%%%
\begin{figure}
\centering
\includegraphics[width=8.9cm]{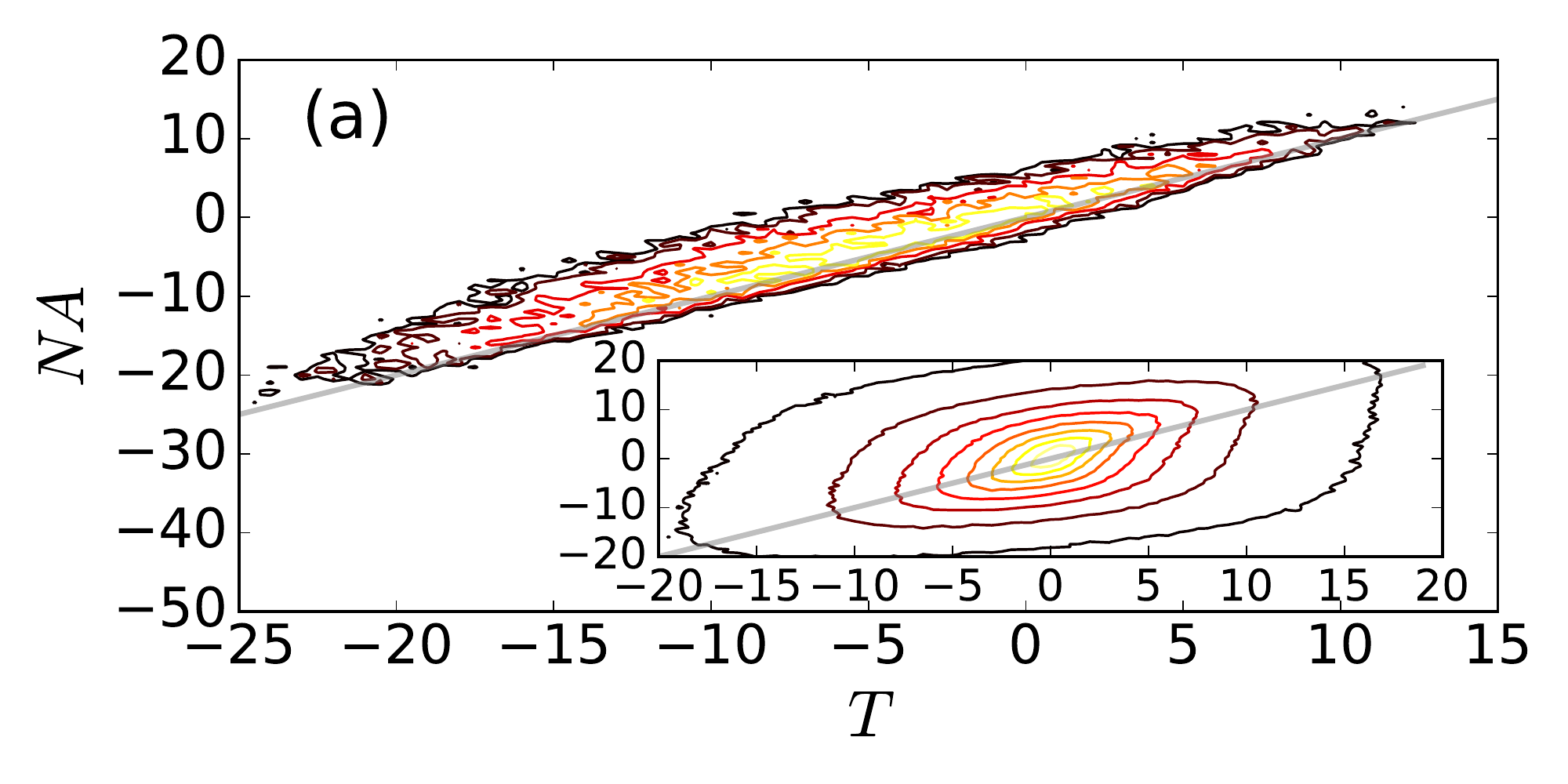}   
\includegraphics[width=8.9cm]{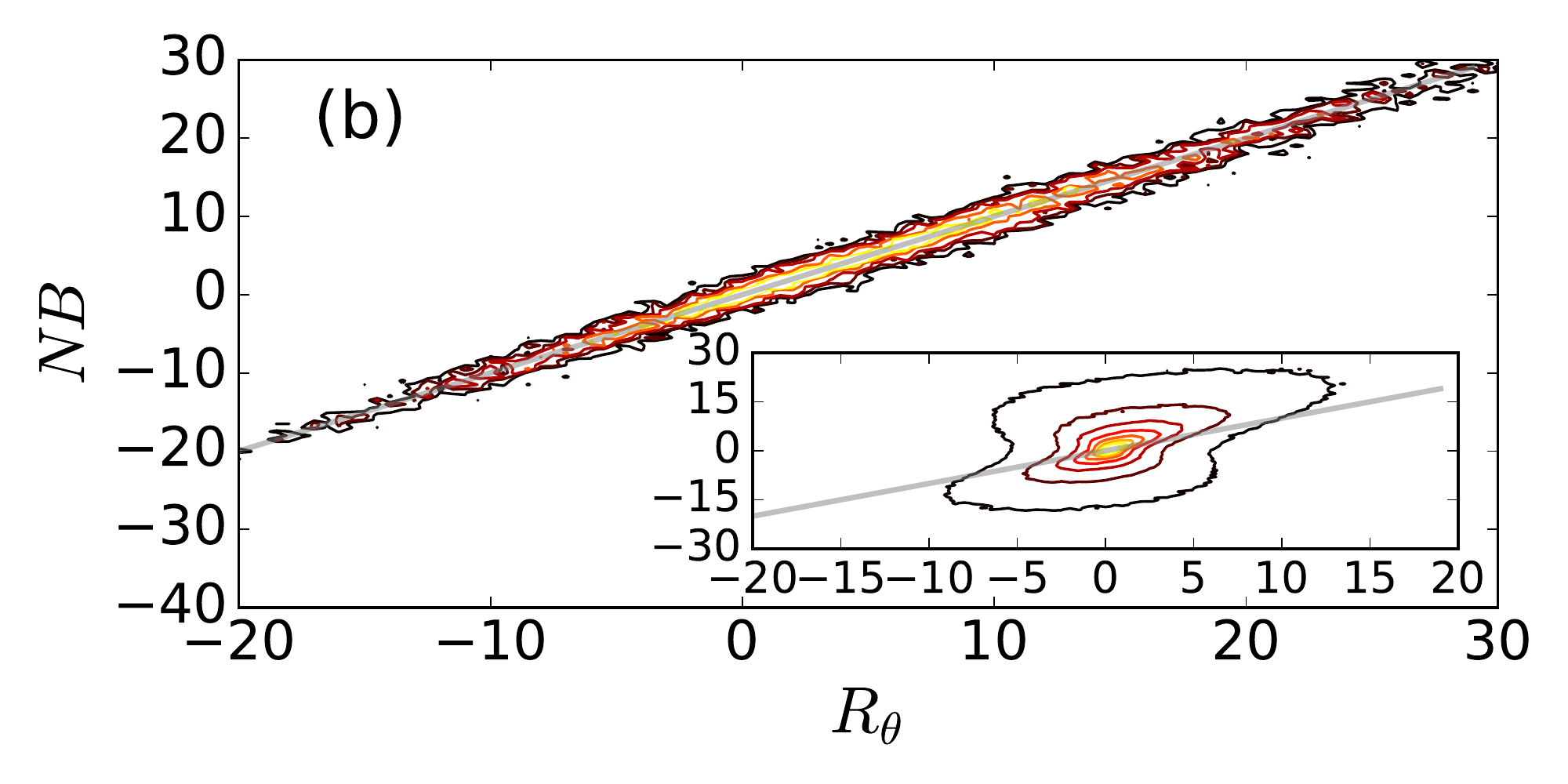}   
\caption{(a) Joint PDF of $T$ and $NA$,
  restricted to particles with $S\approx N$ in the DNS. The gray line
  is $T=NA$. (b) Same for $R_{\theta}$ and $NB$, restricted to
  the same particles. The gray line is $R_{\theta}=NB$. Insets
  show the corresponding joint PDF without any restriction.}
\label{f:AT_RTNB_8}
\end{figure}
%%%%%%%%%%%%%%%%%%%%%%%%%%%%%%%%%%%%%%%%%%

Figure \ref{f:RQ_8} shows the isocontours of the joint probability
density function (PDF) of $Q$ and $R$. As already mentioned, for $N=0$
the Vieillefosse tail $Q+(27R^{2}/4)^{1/3}=0$ is an invariant manifold. The
reduced system blows up following this manifold, with gradients
growing to arbitrarily large 
(negative) values of $Q$. For $N\neq 0$ some fluid elements still
accumulate near this manifold, although accumulation decreases as
$N$ increases (not shown). Figure \ref{f:RQ_8} also shows the joint
PDF of $R_{\theta}$ and $Q$, and as a reference, the relation
$R_{\theta}=(2N^{2}Q-6Q^{2})/(3N)$ (fixed point
$\RN{2}$). There is a correlation between points in the DNS and this
relation (which increases with increasing $N$). The other
lobes in the isocontours are associated to other invariant manifolds of
the ODEs in Eqs.~(\ref{eq:ODEs}) as will be shown next (note these
PDFs correspond to projections in planes of a system with a 7
dimensional phase space).

%%%%%%%%%%%%%%%%%%%%%%%%%%%%%%%%%%%%%%%%%%
%% 		FIGURE 4		%%
%%%%%%%%%%%%%%%%%%%%%%%%%%%%%%%%%%%%%%%%%%
\begin{figure}
\includegraphics[width=8.9cm]{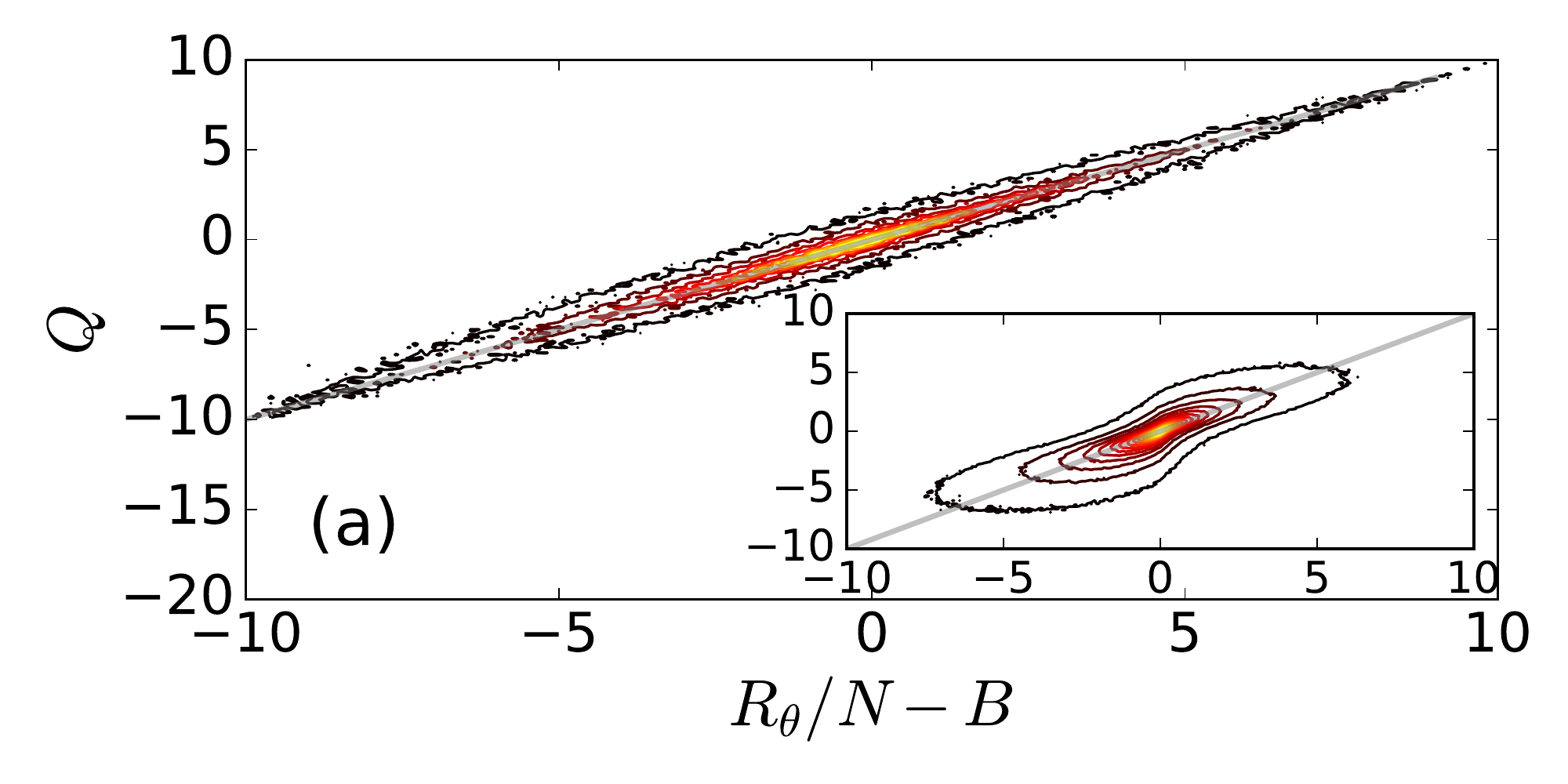}
\includegraphics[width=8.9cm]{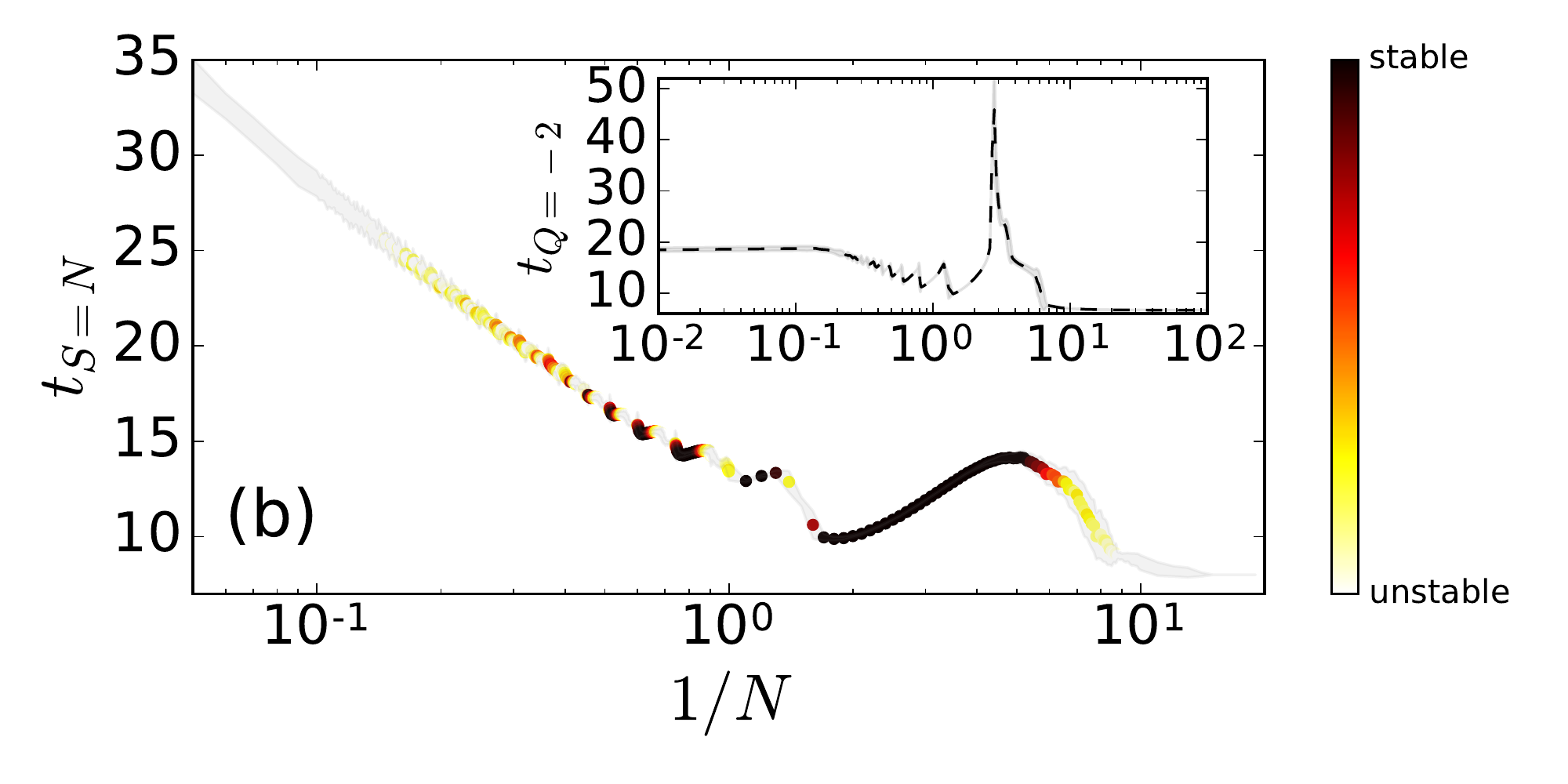}   
\caption{(a) Joint PDF of $R_{\theta}/N-B$ and $Q$, 
  restricted to particles with $S\approx 0$. The inset shows the same
  PDF without restriction. The gray straight lines indicate the manifold
  $Q=R_{\theta}/N-B$. 
  (b) Time $t_{S=N}$ for an ensemble of Eqs.~(\ref{eq:ODEs}) to reach
  $S=N$, from initial values ${\cal O}(10^{-3}-10^{-2})$. Colors
  indicate whether the solution oscillates at $t_{S=N}$, or is already
  growing towards a blow up. The inset shows the time $t_{Q=-2}$ for
  the same ensemble to reach the value $Q=-2$.}
\label{f:RT-B_Q_8}
\end{figure}
%%%%%%%%%%%%%%%%%%%%%%%%%%%%%%%%%%%%%%%%%%

Figure \ref{f:AT_RTNB_8} shows the joint PDFs of $T$ and $NA$, and of
$R_\theta$ and $NB$, obtained from the DNS. As references, we 
show the relations $T=NA$ and $R_{\theta}=NB$ of the central manifold
$\Sigma_0$. The correlation of points in the DNS with this
manifold improves when fluid elements are restricted to the
times when, for each element, $N \approx S$ (as expected for $\Sigma_0$).
Note $\Sigma_0$ corresponds to points with 
$\textrm{Ri}_g \approx 0$, and at the brink of local convection. A
slow evolution, and an accumulation near this manifold, can be
explained as the evolution of the instability (which
takes place in the order of the turnover time) is much slower than
fast internal gravity waves. We can quantify the strength of the
linear relation between two variables $X$ and $Y$ using the Pearson
correlation coefficient, 
$p_{X,Y} = \textrm{cov}(X,Y)/(\sigma_{X} \sigma_{Y})$, 
where cov is the covariance, and $\sigma_X$ and $\sigma_Y$ are the
standard deviations. For $T$ and $NA$ in the runs with
TG forcing, and for fluid elements with $N \approx S$, 
$p_{T,NA} \approx 0.43$ to $0.66$ as $N$ increases. 
%\DEL{$0.61$, and $0.66$, respectively for $N=4$, 8, and 12} 
For RND forcing $p_{T,NA}$ decreases 
%\DEL{e.g.,} 
($p_{T,NA} \approx 0.34$ for $N=8$), which can be expected as this
forcing excites more waves and a flow more stable to local
convection. The same is observed for $R_\theta$ and $NB$; in the runs
with TG forcing and for fluid elements with $N \approx S$,
$p_{R_\theta,NB} \approx 0.85 - 0.96$. In all cases the correlation
increases with increasing stratification; nevertheless, all
simulations show medium to high correlation. Moreover, similar
  behavior is obtained when we condition variables instead on events 
  with $T\approx NA$ or $R_\theta \approx NB$.

As expected, we do not observe correlations associated to 
manifolds $\Sigma_{\RN{2},a}$ and $\Sigma_{\RN{2},b}$. But all
DNSs display points in the vicinity of
$\Sigma_{\RN{1}}$. Figure \ref{f:RT-B_Q_8}(a) shows the joint PDF of 
$R_{\theta}/N-B$ and $Q$, restricted (or not) to fluid elements in the
DNS with $S\approx 0$. Note the strong accumulation in the vicinity
of $R_{\theta}/N-B=Q$, even for large values of $Q$ and of
$R_{\theta}/N-B$ (i.e., away from fixed point $\RN{1}$). For the DNSs
with TG forcing and $N=8$, the Pearson coefficient restricted to
points with $S\approx 0$ is $p_{R_{\theta}/N-B,Q} \approx 0.92$, and
in all DNSs it takes values $\approx 0.81-0.95$ (increasing with
$N$).

%\ADDA{
  In all cases we observe strongerer correlations when statistics
  are restricted to fluid elements with $S= \partial_{z} \theta
  \approx 0$ or $S \approx N$. Table \ref{tab:rest} lists the fraction
  of particles that meet these conditions (within 10\% of the value of
  $N$), for all TG simulations with $768\times768\times192$ grid
  points (and $\textrm{Re} \approx 10^4$). Similar results were
  obtained for the simulation at higher $\textrm{Re}$ and for the RND
  simulation. The table also lists the r.m.s.~value of $\textrm{Ri}_g$
  (averaged over the entire domain), as well as the r.m.s.~value of
  $\textrm{Ri}_g$ for fluid elements with $S\approx 0$ (for fluid
  elements with $S\approx N$, $\textrm{Ri}_g \approx 0$).

\begin{table}[b]
\centering
\setlength{\tabcolsep}{.5 cm}
\begin{tabular*}{\textwidth}{p{1.0cm} c c c c c }
\hline \hline
Forcing & N & Fraction for $S\approx N$ [$\%$] &
  Fraction for $S\approx0$ [$\%$] &
  $\langle Ri_{g}^{2}\rangle^{1/2}$ & 
  $\langle Ri_{g}^{2}\rangle^{1/2}$ for $S\approx0$ \\
\hline
TG & 4  & 8 & 6  & 1900  & 400  \\
TG & 8  & 6 & 9  & 2800  & 1800 \\
TG & 12 & 3 & 17 & 11600 & 24800\\
\hline
\end{tabular*}
\caption{For simulations with TG forcing and $768\times768\times192$
  grid points, and for different Brunt-V\"{a}is\"{a}l\"{a} frequencies
  $N$, the table lists the fraction of fluid elements with 
  $S\approx N$ (in percentage of the total), the fraction of fluid 
  elements with $S\approx 0$ (in percentage), the r.m.s.~value of the
  gradient Richardson number $\textrm{Ri}_g$, and the r.m.s.~value of
  the gradient Richardson number $\textrm{Ri}_g$ restricted to fluid
  elements with $S\approx 0$.}
\label{tab:rest}
\end{table}

%%%%%%%%%%%%%%%%%%%%%%%%%%%%%%%%%%%%%%%%%%%%
%%\section{Blow up}%%
%%%%%%%%%%%%%%%%%%%%%%%%%%%%%%%%%%%%%%%%%%%%

As already mentioned, the ODEs in Eqs.~(\ref{eq:ODEs}) blow up in
finite time, as they amplify gradients nonlinearly. However, although
accumulation along the Vieillefosse tail decreases with increasing
$N$, the blow up time does not increase monotonously with $N$. This
could explain observations of burstiness in stratified turbulence, and
of intermittency in ocean models \cite{pearson_log-normal_2018,
  rorai_turbulence_2014, feraco_vertical_2018}. To finish the
study of the reduced model, we analyze the system stability  
as we vary $1/N$ ($\propto$ Fr). We
integrate an ensemble of Eqs.~(\ref{eq:ODEs}), with $10^{3}$ elements
for each value of $N$, and with initial values $Q=-10^{-2}$,
$R=10^{-2}$, and all other variables with uniformly distributed
random values between $-10^{-3}$ and $10^{-3}$. Figure
\ref{f:RT-B_Q_8}(b) shows the time $t_{S=N}$ it takes the 
ODEs to reach the value $S=N$ as a function of $1/N$ (i.e., the value
for which the ODEs reduce to the central manifold $\Sigma_0$, and
the PDEs can develop overturning instabilities). In
Fig.~\ref{f:RT-B_Q_8}(b) a point is defined as stable when 
$S=N$ was reached with oscillations and not as an already
diverging solution. The inset in Fig.~\ref{f:RT-B_Q_8}(b) shows
the time $t_{Q=-2}$ it takes for the system to reach the arbitrary
value $Q=-2$. As the growth of negative values of $Q$ results in a
blow up, this time also gives a qualitative estimation
of its stability. For large values of $1/N$ (weak stratification) 
the system blows up quickly. As $1/N$ is
decreased and stratification increased, the system becomes at first
more stable, and it takes longer times to reach both $S=N$ and 
$Q=-2$. However, for $1/N$ close to 1, the system alternates between
stable and unstable behavior, and between shorter and
longer times to reach $S=N$. This non-monotonic nonlinear generation
of extreme gradients is compatible with observations in stably
stratified flows \cite{mashayek, rorai_turbulence_2014,
  feraco_vertical_2018} and ocean models
\cite{pearson_log-normal_2018}.

%%%%%%%%%%%%%%%%%%%%%%%%%%%%%%%%%%
\section{Discussion}
%%%%%%%%%%%%%%%%%%%%%%%%%%%%%%%%%%

For a system relevant for atmospheric and oceanic flows with
a huge number of d.o.f., the reduced model with 7 independent
variables can capture many non-trivial properties observed 
in DNSs. The observation that fluid elements evolve in the
vicinity of central invariant manifolds of this system
has several implications:
%(1) As accumulation is observed only in these regions, we are
%confident the reduced model captures the most important
%low-dimensional manifolds of stratified turbulent flows. 
(1) Strong correlation takes place in a manifold that is at the
brink of the convective instability, bringing information on
dynamical properties of these flows. 
(2) New balance relations can be derived from these low-dimensional
manifolds.
%(3) For turbulent flows that are highly anisotropic, the model results
%in a drastic reduction of the number of variables involved in
%prescribing the dynamics in the surroundings of an observer.
(3) The reduced model opens the door to studies of alignment
between temperature and velocity gradients.
%(asthe independent variables are scalar reductions of these field
%gradients), and, for short times (as we neglect the pressure Hessian)
%to predict the evolution of gradients in the Lagrangian framework,
%useful for observations in the atmosphere and the ocean.
(4) Finally, the non-monotonicity of the reduced system with the level
of stratification can explain recent observations of enhanced extreme
events in stably stratified turbulence.
%As a result, the system of ODEs presented here opens a new path to
%the usage  of restricted models to study complex and realistic flows.

%The reduced model derived here has several implications for
%atmospheric and oceanic flows.
  As an example of these implications, the potential vorticity (an
  important quantity to understand motions in geophysical flows, that
  can only change due to external forcing or dissipation) for this
  system can be written as 
  $P_V = \bom \cdot \nabla \theta - N \omega_z 
    = \theta_x (A_{zy} - A_{yz}) + \theta_y (A_{xz} - A_{zx}) +
    (S-N) (A_{yx} - A_{xy})$, where $\bom$ is the vorticity. 
  The third term vanishes for fluid elements in $\Sigma_0$, and our
  DNSs confirm that those points take smaller values of $P_V$. As a
  second example, many quantities in the ODEs are associated to
  production of small scales by turbulence, as, e.g., 
  $T=\partial_i \, \theta \partial_z u_i$ which quantifies nonlinear 
  (stretching) production of small-scale vertical temperature
  fluctuations \cite{gulitski_2007}. From our results, for fluid
  elements in $\Sigma_0$ the turbulent production $T$ is balanced by
  the linear (buoyancy) production $NA$. Finally, as the reduced model
  provides information on production of strain and of temperature
  gradients, it can be used to design sub-grid scale parameterizations
  for mixing and dissipation in atmospheric and oceanic models that
  can better capture the observed small-scale intermittency
  \cite{pearson_log-normal_2018}, following procedures previously 
  used for reduced models of HIT \cite{meneveau_lagrangian_2011}. As a
  result, the system of ODEs presented here opens a new path to the
  usage of restricted models to study complex and anisotropic flows.

\begin{acknowledgments}
{\it NES and PDM acknowledge support from PICT Grant No.~2015-3530.}
\end{acknowledgments}

\bibliography{ms}

\end{document}